\documentclass[11pt,a4paper]{article}
\usepackage[utf8]{inputenc}
\usepackage[T1]{fontenc}
\usepackage{lmodern}
\usepackage{amsmath,amssymb,amsthm}
\usepackage[margin=2.6cm,marginparwidth=2.0cm]{geometry}
\usepackage{xcolor}
\usepackage{microtype}
\usepackage[colorlinks=true,linkcolor=black,citecolor=black,urlcolor=black]{hyperref}

\newcommand{\Tr}{\operatorname{Tr}}
\newcommand{\ad}{\operatorname{ad}}
\newcommand{\Ad}{\operatorname{Ad}}
\newcommand{\vd}[2]{\frac{\delta #1}{\delta #2}}
\newtheorem{definition}{Definition}[section]
\theoremstyle{remark}
\newtheorem{remark}[definition]{Remark}

\title{Euler--Poincar\'e Reduction: A Geometric Treatment of the Relativistic Spinning Particle}
\author{Burak G\"ul\thanks{Elite Master's Program in Theoretical and Mathematical Physics (TMP), Ludwig-Maximilians-Universit\"at M\"unchen \& Technische Universit\"at M\"unchen, Munich, Germany}
\and Osman Teoman Turgut\thanks{Department of Physics, Bo\u{g}azi\c{c}i University, Istanbul, Turkey}}
\date{}

\begin{document}
\maketitle

\begin{abstract}
The heavy top is a classic example of Euler--Poincar\'e reduction on a semidirect product group, in which an advected quantity---the direction of gravity---couples to rotational motion. Relativistic spinning particles whose configuration space is the Poincar\'e group are well understood in the Hamiltonian, symplectic and constrained-dynamics frameworks. Here we give an explicit Lagrangian Euler--Poincar\'e treatment of such a particle in left-trivialized (body) coordinates, starting from the Lagrangian of Skagerstam and Stern. For the free particle the reduction is complete, and the computation runs parallel to that of the heavy top, with the vector-space factor $\mathbb{R}^{3,1}$ of the Poincar\'e group taking the place of the space of gravity directions. We show that this parallel is structural but not dynamical: unlike the direction of gravity, the spacetime position is a dynamical variable and is not advected. Coupling to an external electromagnetic field reintroduces an explicit dependence on the group element through the fields, so that only a partial reduction is possible; the resulting equations give the generalized force and torque equations, which reduce to the Bargmann--Michel--Telegdi equation for weak homogeneous fields. Finally, we observe that in left-trivialized coordinates the spin--field coupling of Skagerstam and Stern appears as a field-dependent redefinition of the particle's mass, which makes this coupling particularly convenient within the Euler--Poincar\'e framework.
\end{abstract}

\noindent{\raggedright\textbf{Keywords:} Euler--Poincar\'e reduction; semidirect products; relativistic spinning particle; Poincar\'e group; heavy top; Bargmann--Michel--Telegdi equation.\par}

\section{Introduction}

The formulation of a consistent classical mechanics for relativistic particles with intrinsic spin has a rich history in theoretical physics. While the translational dynamics of point particles are straightforwardly described by trajectories in Minkowski spacetime, incorporating spin requires extending the configuration space to capture internal rotational degrees of freedom. Geometrically, this is naturally achieved by identifying the particle's configuration space with the Poincar\'e group. This perspective is well established in the Hamiltonian and symplectic frameworks; Souriau famously demonstrated that elementary particles with spin can be rigorously classified as coadjoint orbits of the Poincar\'e group~\cite{Souriau}. Building on this symmetry-driven approach, Hanson and Regge~\cite{HansonRegge} provided a comprehensive canonical treatment of the relativistic spherical top, analyzing its constrained phase space via Dirac brackets.

Lagrangians defined directly on the Poincar\'e group were developed by Balachandran, Marmo, Skagerstam and Stern~\cite{BMSS80,BMSS83} and by Skagerstam and Stern~\cite{SkagerstamStern}, who coupled the particle to external electromagnetic and gravitational fields and recovered the Bargmann--Michel--Telegdi (BMT)~\cite{BMT} and Mathisson--Papapetrou equations. Cho, Hyun and Kim~\cite{CHK94} emphasized that such Lagrangians are built from the two left-invariant one-forms $\Lambda^{-1}d\Lambda$ and $\Lambda^{-1}dz$ on the Poincar\'e group and analysed the resulting constraints. K\"unzle~\cite{Kunzle} constructed the canonical dynamics of spinning particles in gravitational and electromagnetic fields, and Bailey and Israel~\cite{BaileyIsrael} gave a Lagrangian treatment in general relativity. More recently, the constrained Hamiltonian dynamics on the Poincar\'e group viewed as a configuration space was analysed in detail by Andrzejewski et al.~\cite{AGGKM22}, who showed that the first-class constraints generate the stabilizer of a point on the coadjoint orbit; the spherical top has been related to massive twistors~\cite{KKL21}, and several spinning-particle models, including the spherical top, have been shown to describe the same physics both free and coupled to electromagnetism and gravity~\cite{KL26}. Variational formulations of the Frenkel and BMT equations with other choices of spin variables were given in~\cite{Deriglazov12,DeriglazovPM14}; see also~\cite{Zakrzewski95,LSS96,Steinhoff15} for related constructions.

On the Lagrangian side of geometric mechanics, the systematic tool for systems on Lie groups is Euler--Poincar\'e reduction~\cite{MarsdenRatiu,HolmSchmahStoica}. Its semidirect-product version, developed by Holm, Marsden and Ratiu~\cite{HMR98} and by Cendra, Holm, Marsden and Ratiu~\cite{CHMR98}, treats systems such as the heavy top, in which a parameter (the direction of gravity) breaks the symmetry and is restored as an advected quantity. To our knowledge, the relativistic spinning particle has not been treated explicitly within this Lagrangian reduction framework, and this is the purpose of the present paper.

Our contributions are as follows. (i) We carry out the Euler--Poincar\'e reduction of the free spinning particle of~\cite{SkagerstamStern} in left-trivialized coordinates and show that the computation runs parallel to that of the heavy top: in both cases one works on a semidirect product $G\ltimes V$ and with the body representation $\Gamma=g^{-1}v$ of a vector in $V$. We also state the limits of this parallel: the spacetime position is a dynamical variable rather than an advected parameter, and the free Lagrangian is in fact invariant under the full Poincar\'e group, which connects the reduced equations to the coadjoint-orbit description. (ii) We show that coupling to an external electromagnetic field reintroduces an explicit dependence on the group element through the fields, so that only a partial reduction is possible, and we derive the corresponding force and torque equations, recovering the BMT equation for weak homogeneous fields. (iii) We observe that in left-trivialized coordinates the spin--field coupling of~\cite{SkagerstamStern} appears as a field-dependent redefinition of the mass, which makes it more convenient in this framework than the alternative coupling discussed there. Throughout, we keep the presentation self-contained and use the heavy top as a guide, so that the paper can also serve as an introduction to these methods.

The paper is organized as follows. In Section~\ref{sec:prelim}, we introduce the necessary mathematical preliminaries, establishing the notation for Lie groups, dual algebras, and coadjoint actions. Section~\ref{sec:EP} details the general procedure of Euler--Poincar\'e reduction. To build physical intuition, we first review the free rigid body and the heavy top, highlighting how the semidirect product structure accommodates advected quantities. We then apply the same framework to the free relativistic spinning particle, making the structural parallel with the heavy top explicit. Finally, we introduce external electromagnetic fields, showing how the coordinate dependence of the fields breaks full symmetry and leads to a partial reduction only. From this, we derive the generalized force and torque equations, recovering the BMT equation under weak-field approximations. Section~\ref{sec:concl} summarizes our results and discusses open directions. Detailed variational calculations for the interacting Lagrangian are provided in the Appendix.

\section{Mathematical Preliminaries}\label{sec:prelim}

We begin by introducing the relevant mathematical background and reviewing the method of Euler--Poincar\'e reduction for a general Lie group; specializing to $SO(3)$ gives the familiar rigid body.

Let $G$ be a Lie group and $\mathfrak g=T_eG$ its Lie algebra. The left and right translations by a group element $g\in G$ are denoted by $L_g:h\mapsto gh$ and $R_g:h\mapsto hg$, respectively. The duality pairing between the Lie algebra $\mathfrak g$ and its dual $\mathfrak g^*$ is denoted by
\begin{equation}
\langle\mu,\xi\rangle:\mathfrak g^*\times\mathfrak g\to\mathbb R,
\end{equation}
which is a non-degenerate bilinear form. For $\mu\in\mathfrak g^*$ and $\xi\in\mathfrak g$, this pairing represents the action of the covector $\mu$ on the vector $\xi$.

The adjoint action of $G$ on $\mathfrak g$ is $\Ad_g:\mathfrak g\to\mathfrak g$, defined by $\Ad_g\xi=T_e(R_{g^{-1}}\circ L_g)\xi$. The adjoint action of $\mathfrak g$ on itself is the Lie bracket, $\ad_\xi\eta=[\xi,\eta]$. The coadjoint actions are induced actions on the dual space $\mathfrak g^*$, denoted by $\Ad^*_{g^{-1}}$ and $\ad^*_\xi$, satisfying for all $\mu\in\mathfrak g^*$ and $\xi,\eta\in\mathfrak g$:
\begin{equation}
\langle\ad^*_\xi\mu,\eta\rangle=\langle\mu,\ad_\xi\eta\rangle=\langle\mu,[\xi,\eta]\rangle.
\end{equation}

For the Lie group $SO(3)$, its Lie algebra $\mathfrak{so}(3)$ consists of $3\times3$ skew-symmetric matrices. We identify $\mathfrak{so}(3)$ with $\mathbb R^3$ via the following isomorphisms:

\emph{Hat map} $\hat{(\cdot)}:\mathbb R^3\to\mathfrak{so}(3)$, defined for $\omega=(\omega_1,\omega_2,\omega_3)^T\in\mathbb R^3$ by
\begin{equation}
\hat\omega=\begin{pmatrix}0&-\omega_3&\omega_2\\ \omega_3&0&-\omega_1\\ -\omega_2&\omega_1&0\end{pmatrix}\in\mathfrak{so}(3).
\end{equation}
This map satisfies $\hat\omega v=\omega\times v$ for any $v\in\mathbb R^3$.

\emph{Breve map} $\breve{(\cdot)}:\mathbb R^3\to\mathfrak{so}(3)^*$, identifying the dual algebra with $\mathbb R^3$. Under this identification, the duality pairing becomes the standard dot product:
\begin{equation}
\langle\breve\mu,\hat\omega\rangle=\mu\cdot\omega,
\end{equation}
where $\mu\in\mathbb R^3$ corresponds to an element of $\mathfrak{so}(3)^*$ via the breve map. With these identifications, the coadjoint action for $\mathfrak{so}(3)$ becomes
\begin{equation}
\ad^*_{\hat\omega}\breve\mu=\mu\times\omega,
\end{equation}
which follows from the property $[\hat\omega,\hat\eta]=\widehat{\omega\times\eta}$.

For Minkowski spacetime $\mathbb R^{3,1}$ we use the metric $\eta=\mathrm{diag}(-1,1,1,1)$ throughout; indices are raised and lowered with $\eta$.

\section{Euler--Poincar\'e Reduction for a Lie Group}\label{sec:EP}

In geometric mechanics, systems with continuous symmetries can often be simplified by a process called \emph{reduction}. When the configuration space is a Lie group $G$ and the Lagrangian is $G$-invariant, the original Euler--Lagrange equations on $TG$ can be reduced to a new set of equations on the Lie algebra $\mathfrak g=T_eG$, or its dual $\mathfrak g^*$. These are the \emph{Euler--Poincar\'e equations}~\cite{HolmSchmahStoica,MarsdenRatiu}.

This section provides a detailed exposition of Euler--Poincar\'e reduction for the convenience of the reader. It closely follows the conceptual and notational style found in the works of Darryl D.\ Holm, particularly his books and lecture notes on geometric mechanics~\cite{HolmSchmahStoica}, and the semidirect-product theory of~\cite{HMR98,CHMR98}.

Consider a Lagrangian $L:TG\to\mathbb R$ that may depend explicitly on the group element $g$. We define the \emph{body representation} of the velocity:
\begin{equation}
\xi=g^{-1}\dot g\in\mathfrak g,
\end{equation}
and express the Lagrangian in terms of the so-called left-trivialized (body) coordinates $(g,\xi)\in G\times\mathfrak g$ as:
\begin{equation}
\tilde L(g,\xi):=L(g,g\xi)=L(g,\dot g).
\end{equation}

\subsection{Variational Principle}
The action functional $S=\int_{t_1}^{t_2}\tilde L(g,\xi)\,dt$ is stationary under variations $\delta g$ that vanish at the endpoints. Let $\delta g(t)=\frac{d}{d\epsilon}\big|_{\epsilon=0}g_\epsilon(t)$ and define the body variation $\eta(t)\in\mathfrak g$ by
\begin{equation}
\eta(t)=g(t)^{-1}\delta g(t),
\end{equation}
which also vanishes at the endpoints. The variation of $\xi$ can be calculated as follows: using $\delta(g^{-1})=-g^{-1}\delta g\,g^{-1}$ and $\frac{d}{dt}(g^{-1})=-g^{-1}\dot g\,g^{-1}$,
\begin{equation}
\delta\xi=-g^{-1}\delta g\,g^{-1}\dot g+g^{-1}\delta\dot g=-\eta\xi+g^{-1}\delta\dot g,\qquad
\dot\eta=-g^{-1}\dot g\,g^{-1}\delta g+g^{-1}\delta\dot g=-\xi\eta+g^{-1}\delta\dot g,
\end{equation}
so that, subtracting,
\begin{equation}
\delta\xi=\dot\eta+[\xi,\eta]=\dot\eta+\ad_\xi\eta.
\end{equation}
In these coordinates, the variation of the action is:
\begin{align}
\delta S&=\int_{t_1}^{t_2}\Big[\Big\langle\vd{\tilde L}{g},\delta g\Big\rangle+\Big\langle\vd{\tilde L}{\xi},\delta\xi\Big\rangle\Big]dt\\
&=\int_{t_1}^{t_2}\Big[\Big\langle\vd{\tilde L}{g},g\eta\Big\rangle+\Big\langle\vd{\tilde L}{\xi},\dot\eta+\ad_\xi\eta\Big\rangle\Big]dt.
\end{align}
We integrate the second term by parts:
\[
\int_{t_1}^{t_2}\Big\langle\vd{\tilde L}{\xi},\dot\eta\Big\rangle dt=\Big[\Big\langle\vd{\tilde L}{\xi},\eta\Big\rangle\Big]_{t_1}^{t_2}-\int_{t_1}^{t_2}\Big\langle\frac{d}{dt}\vd{\tilde L}{\xi},\eta\Big\rangle dt,
\]
and the boundary term vanishes since $\eta(t_1)=\eta(t_2)=0$.

Now, we need to express $\langle\delta\tilde L/\delta g,g\eta\rangle$ in terms of $\eta$. Observe that $g\eta=T_eL_g(\eta)$, since left translation maps the Lie algebra element $\eta$ to a tangent vector at $g$. Using the duality pairing, we have:
\begin{equation}
\Big\langle\vd{\tilde L}{g},g\eta\Big\rangle=\Big\langle\vd{\tilde L}{g},T_eL_g(\eta)\Big\rangle=\Big\langle T_e^*L_g\Big(\vd{\tilde L}{g}\Big),\eta\Big\rangle\qquad\forall\eta\in\mathfrak g,
\end{equation}
where $T_e^*L_g:T_g^*G\to\mathfrak g^*$ is the dual of the tangent map $T_eL_g$. This defines an element of $\mathfrak g^*$ that encodes the explicit $g$-dependence of the Lagrangian. With this expression, the variation becomes:
\[
\delta S=\int_{t_1}^{t_2}\Big\langle-\frac{d}{dt}\vd{\tilde L}{\xi}+\ad^*_\xi\vd{\tilde L}{\xi}+T_e^*L_g\Big(\vd{\tilde L}{g}\Big),\eta\Big\rangle dt=0.
\]
Since $\eta$ is arbitrary, we obtain the general Euler--Poincar\'e equation~\cite{HolmSchmahStoica}:
\begin{equation}\label{eq:EPgeneral}
\frac{d}{dt}\vd{\tilde L}{\xi}=\ad^*_\xi\vd{\tilde L}{\xi}+T_e^*L_g\Big(\vd{\tilde L}{g}\Big).
\end{equation}
In case the Lagrangian is left invariant, then:
\begin{equation}
\tilde L(g,\xi)=L(g,\dot g)=L(g^{-1}g,g^{-1}\dot g)=L(e,\xi),
\end{equation}
and we define the reduced Lagrangian to be the restriction of $L$ to $\mathfrak g$,
\begin{equation}
l:\mathfrak g\to\mathbb R,\qquad l(\xi):=\tilde L(g,\xi)=L(e,\xi).
\end{equation}
For a left-invariant Lagrangian, $\tilde L$ does not depend explicitly on $g$, so $\delta\tilde L/\delta g=0$, hence the extra term vanishes, and~\eqref{eq:EPgeneral} becomes
\begin{equation}\label{eq:EP}
\frac{d}{dt}\vd{l}{\xi}=\ad^*_\xi\vd{l}{\xi}.
\end{equation}
This process of using left invariance to restrict the domain of the Lagrangian from $TG$ to $\mathfrak g$ is called \emph{Euler--Poincar\'e reduction}.

\subsection{Example 1: The Free Rigid Body}
For $G=SO(3)$, we define $\hat\Omega=R^{-1}\dot R\in\mathfrak{so}(3)$. The Lagrangian for the free rigid body is just the kinetic energy, and it is left invariant:
\begin{equation}
L(R,\dot R)=\frac12\Tr\big(\dot R\,\mathbb J\,\dot R^T\big)=\frac12\,\Omega\cdot\mathbb I\,\Omega=l(\Omega),
\end{equation}
where $\mathbb I=\int\rho(x)\big(\|x\|^2\,\mathbf 1-xx^T\big)d^3x$ is the moment of inertia tensor and $\mathbb J=\int\rho(x)\,xx^T d^3x$ is its coefficient matrix, both integrated over the body. Then we can apply~\eqref{eq:EP} for the reduced Lagrangian.

To put this equation into a more familiar form, we can use the hat and breve isomorphisms, identifying $\hat\Omega$ with the body angular velocity $\Omega\in\mathbb R^3$ and $\breve\Pi:=\delta l/\delta\hat\Omega\in\mathfrak{so}(3)^*$ with $\Pi=\delta l/\delta\Omega\in\mathbb R^3$. The variational derivative is the angular momentum:
\begin{equation}
\Pi=\vd{l}{\Omega}=\mathbb I\,\Omega\in\mathbb R^3.
\end{equation}
Using the vector representation~(5) for the coadjoint action, $\ad^*_\Omega\Pi=\Pi\times\Omega$, \eqref{eq:EP} becomes:
\begin{equation}
\frac{d}{dt}\Pi=\Pi\times\Omega.
\end{equation}
This is the well-known Euler's equations for a free rigid body~\cite{HolmSchmahStoica,MarsdenRatiu}.

\subsection{Example 2: The Heavy Top}\label{sec:heavytop}
In the case of an external potential that depends on $g\in G$, we might lose the initial $G$-invariance of our Lagrangian. In such cases, it is possible to extend the configuration space of our Lagrangian to obtain an extended Lagrangian that is invariant under the initial symmetry. We can then use this symmetry to eliminate the $g$ dependence of the extended Lagrangian and apply the reduction process.

The standard example is the heavy top, which is a rigid body with a fixed point in a gravitational field, in the direction of the unit vector $k$. Let $\chi\in\mathbb R^3$ be the fixed vector from the support to the body's centre of mass. The potential
\begin{equation}
V_k(R)=mg\langle k,R\chi\rangle
\end{equation}
breaks the full rotational symmetry of the Lagrangian. To recover a full symmetry, we extend the configuration space to the semidirect product $G=SO(3)\ltimes\mathbb R^3$, so that the direction of gravity $k$ can be considered as a value of the new coordinate $v\in(\mathbb R^3)^*$. So we can write the extended Lagrangian as
\begin{equation}
L:TSO(3)\times(\mathbb R^3)^*\to\mathbb R,\qquad L(R,\dot R,v)=\frac12\Tr\big(\dot R\,\mathbb J\,\dot R^T\big)-mg\langle v,R\chi\rangle,
\end{equation}
with an additional constraint
\begin{equation}
v(t)=k.
\end{equation}

\begin{definition}[Diagonal action]
The (left) diagonal action of $SO(3)$ on $SO(3)\times(\mathbb R^3)^*$ is:
\begin{equation}
(Q,(R,v))\mapsto(QR,Qv),\qquad\text{for all }Q\in SO(3).
\end{equation}
Its tangent lift is given by
\begin{equation}
\big(Q,(R,\dot R,v,\dot v)\big)\mapsto\big(QR,Q\dot R,Qv,Q\dot v\big).
\end{equation}
\end{definition}

The Lagrangian above is invariant under this action. Defining the left-trivialized (body) coordinates $\Gamma=R^{-1}v$ and $\hat\Omega=R^{-1}\dot R$, the tangent-lifted left diagonal action can be written in these coordinates as
\begin{equation}
\big(Q,(R,\hat\Omega,\Gamma,\dot\Gamma)\big)\mapsto\big(QR,\hat\Omega,\Gamma,\dot\Gamma\big).
\end{equation}
Choosing $Q=R^{-1}$, we can eliminate the $R$ dependence of the Lagrangian. Indeed, switching to the body coordinates gives the reduced Lagrangian as
\begin{equation}
l(\Omega,\Gamma)=\frac12\langle\Omega,\mathbb I\,\Omega\rangle-mg\langle\Gamma,\chi\rangle.
\end{equation}
The Euler--Poincar\'e equations for this semidirect-product system are~\cite{HMR98,HolmSchmahStoica}
\begin{equation}
\dot\Pi=\Pi\times\Omega+mg\,\Gamma\times\chi,\qquad\Pi=\mathbb I\,\Omega, \tag{27a}
\end{equation}
\begin{equation}
\dot\Gamma=\Gamma\times\Omega. \tag{27b}
\end{equation}
Equation~(27b) is the \emph{advection equation}: it expresses only that $v=k$ is constant in space and that $\Gamma$ is its body representation. In the variational principle, $\Gamma$ carries no independent variation; its variation $\delta\Gamma=-\hat\eta\,\Gamma$ is induced entirely by the variation of $R$~\cite{HMR98,CHMR98}. This is what makes $\Gamma$ an advected quantity, and it is the point at which the comparison with the spinning particle below must be made carefully. The variational principle in these coordinates is done explicitly in the analogous case below for a relativistic particle and will not be repeated here; the details for the heavy top can be found in~\cite{HMR98,HolmSchmahStoica}.

\subsection{Free Spinning Particle}\label{sec:free}
While the description of relativistic spinning particles via Lie group configurations and coadjoint orbit reductions is well known in Hamiltonian mechanics and in the symplectic formalism~\cite{Souriau,HansonRegge,Kunzle,AGGKM22}, and Lagrangians on the Poincar\'e group have been studied before~\cite{BMSS80,BMSS83,SkagerstamStern,CHK94}, an explicit Euler--Poincar\'e reduction of such a Lagrangian in left-trivialized coordinates has not, to our knowledge, appeared in the literature. Carrying it out makes the structural parallel with the heavy top transparent.

We will now apply this reduction process to determine the relativistic equations of motion for a free spinning particle. The configuration space is the Poincar\'e group
\begin{equation}
G=SO(3,1)\ltimes\mathbb R^{3,1},
\end{equation}
where $SO(3,1)$ is the Lorentz group (proper orthochronous) and $\mathbb R^{3,1}$ is Minkowski spacetime. The Lagrangian is defined on the tangent bundle:
\begin{equation}
L_0:TSO(3,1)\times T\mathbb R^{3,1}\to\mathbb R.
\end{equation}
In coordinates, $\Lambda\in SO(3,1)$ represents a Lorentz transformation that relates the body frame to a fixed inertial frame, and $z\in\mathbb R^{3,1}$ represents a spacetime vector. We use the Lagrangian given in~\cite{SkagerstamStern}:
\begin{equation}
L_0=p^{\mu}\dot z_{\mu}+\frac{i\lambda}{2}\Tr\big(M_{12}\Lambda^{-1}\dot\Lambda\big),
\end{equation}
where $p^{\mu}=m\Lambda^\mu{}_0$ is the momentum, $M_{12}$ is a generator of Lorentz transformations given in the body frame (specifically the rotation generator in the $12$-plane), and $\lambda$ is a constant; explicitly, $M^{\mu\nu}=-iK^{\mu\nu}$ with $(K^{\mu\nu})^\rho{}_\sigma=\eta^{\mu\rho}\delta^\nu{}_\sigma-\eta^{\nu\rho}\delta^\mu{}_\sigma$, which is the normalization for which~(38) below holds. These identifications are natural from a physical point of view. A Lagrangian of the same form, built from the left-invariant one-forms $\Lambda^{-1}d\Lambda$ and $\Lambda^{-1}dz$, is also the starting point of~\cite{CHK94}.

We define our body-frame coordinates analogous to the previous case, $\hat\Omega=\Lambda^{-1}\dot\Lambda\in\mathfrak{so}(3,1)$ and $\Gamma_\mu=(\Lambda^{-1}z)_\mu\in\mathbb R^{3,1}$. The Lagrangian~(30) is invariant under the left diagonal action analogous to~(25). This can be seen from
\begin{equation}
\begin{aligned}
m\Lambda^\mu{}_0\dot z_\mu\;\to\;m(Q\Lambda)^\mu{}_0(Q\dot z)_\mu&=mQ^\mu{}_\nu\Lambda^\nu{}_0(Q^{-1})^\rho{}_\mu\dot z_\rho\\
&=m\Lambda^\nu{}_0\dot z_\nu.
\end{aligned}
\end{equation}
Since $L_0$ depends on $z$ only through $\dot z$, it is also invariant under translations $z\mapsto z+a$; hence $L_0$ is invariant under the full left action of the Poincar\'e group. We return to this point in Remark~\ref{rem:full}.
Indeed, in body coordinates the Lagrangian becomes:
\begin{equation}
\tilde L_0(\Lambda,\hat\Omega,\Gamma,\dot\Gamma)=m\dot\Gamma_0-m\hat\Omega^\nu{}_0\Gamma_\nu+\frac{i\lambda}{2}\Tr\big(M_{12}\hat\Omega\big),
\end{equation}
which is independent of $\Lambda$ as expected by the above invariance. The variation in these coordinates can be obtained as follows:
\begin{equation}
\delta\hat\Omega=\dot{\hat\Sigma}+[\hat\Omega,\hat\Sigma]=\dot{\hat\Sigma}+\ad_{\hat\Omega}\hat\Sigma,\qquad
\delta\Gamma=-\hat\Sigma\Gamma+W,
\end{equation}
where $\hat\Sigma=\Lambda^{-1}\delta\Lambda\in\mathfrak{so}(3,1)$ and $W=\Lambda^{-1}\delta z\in\mathbb R^{3,1}$ are arbitrary variational paths, vanishing at the endpoints.

Plugging the above variations into~(32) then yields the following result,
\begin{equation}
\begin{aligned}
\delta\tilde L_0=&\Big(m\dot\Gamma_\alpha-m\Gamma_\beta\hat\Omega^\beta{}_\alpha\delta^\mu{}_0+\frac{i\lambda}{2}(M^{12})^{\mu\nu}\big(\hat\Omega_{\nu\alpha}-\hat\Omega_{\alpha\nu}\big)\Big)\hat\Sigma^\alpha{}_\nu\\
&-\big(m\hat\Omega^\alpha{}_0+\dot m\,\delta^\alpha{}_0\big)W_\alpha.
\end{aligned}
\end{equation}
We set both expressions equal to $0$ separately since the variations are independent. For the first line in~(34), we can expand $\hat\Sigma$ in terms of the generators as
\begin{equation}
\hat\Sigma=i\big(\Lambda^{-1}M^{\alpha\beta}\Lambda\big)\theta_{\alpha\beta}
\end{equation}
(notice that we are not using $M^{\alpha\beta}\theta_{\alpha\beta}$ because the Lorentz algebra generators are defined in the body frame and $\Lambda^{-1}M^{\alpha\beta}\Lambda$ accounts for the frame dependency of $\hat\Sigma:=\Lambda^{-1}\delta\Lambda$). For the second line in~(34), we just use the definition $W=\Lambda^{-1}\delta z$. Plugging these, we obtain
\begin{equation}
\begin{aligned}
0=&\big(\dot S^{\mu\nu}+\dot z^\mu p^\nu-\dot z^\nu p^\mu\big)\theta_{\mu\nu}\\
&+\big(m\dot\Lambda_{\mu0}+\dot m\Lambda_{\mu0}\big)\delta z^\mu,
\end{aligned}
\end{equation}
where
\[
S^{\mu\nu}:=\frac\lambda2\Tr\big(M^{12}\Lambda^{-1}M^{\mu\nu}\Lambda\big)
\]
will be referred to as the spin tensor. Notice that this tensor is orthogonal to the momentum,
\begin{equation}
S^{\mu\nu}p_\mu=0.
\end{equation}
From~(36), we get two equations. The first one is
\begin{equation}
\dot S^{\mu\nu}+\dot z^\mu p^\nu-\dot z^\nu p^\mu=0,
\end{equation}
and the other equation is simply the momentum conservation $\dot p^\mu=0$~\cite{SkagerstamStern}. Together, these state that the total angular momentum
\[
J^{\mu\nu}:=z^\mu p^\nu-z^\nu p^\mu+S^{\mu\nu}
\]
is conserved. Moreover, differentiating the identity~(37) and using~(38) and $\dot p=0$ gives $p^2\dot z^\mu=(p\cdot\dot z)\,p^\mu$, so that $\dot z\parallel p$. Hence $\dot z^\mu p^\nu-\dot z^\nu p^\mu=0$, and the spin tensor and the orbital angular momentum are in fact separately conserved.

\paragraph{The parallel with the heavy top, and its limits.}
The computation above runs parallel to that of the heavy top: in both cases one works on a semidirect product $G\ltimes V$ with $G$ acting on $V$, the Lagrangian is invariant under the left (diagonal) action of $G$, and the vector-space variable enters the reduced Lagrangian through its body representation $\Gamma=g^{-1}v$. The parallel is, however, structural rather than dynamical. For the heavy top, $v=k$ is a fixed parameter: $\Gamma$ has no independent variation, and its evolution~(27b) is a kinematic consequence of $k$ being constant in space---this is what makes it an advected quantity in the sense of~\cite{HMR98,CHMR98}. For the spinning particle, $z$ is a dynamical variable: its body representation $\Gamma=\Lambda^{-1}z$ has the independent variation $W$ in~(33), and this variation produces the additional equation $\dot p=0$, which has no counterpart for the heavy top. Correspondingly, the reduced Lagrangian~(32) contains $\dot\Gamma$ and is not of the heavy-top form~(27), in which $\Gamma$ enters only through a potential. In this sense the spacetime position plays the role of the semidirect-product variable of the heavy top, but it is not an advected quantity.

\begin{remark}[Full Poincar\'e invariance and coadjoint orbits]\label{rem:full}
Since $L_0$ is invariant under the full Poincar\'e group $P=SO(3,1)\ltimes\mathbb R^{3,1}$, one may also reduce directly by $P$. With the group law $(\Lambda,a)(\Lambda',a')=(\Lambda\Lambda',\Lambda a'+a)$, the left-trivialized velocity of $g=(\Lambda,z)$ is
\begin{equation}
\xi:=g^{-1}\dot g=\big(\Lambda^{-1}\dot\Lambda,\;\Lambda^{-1}\dot z\big)=(\hat\Omega,u)\in\mathfrak p, \tag{38a}
\end{equation}
and, since $\dot\Gamma=\frac{d}{dt}(\Lambda^{-1}z)=-\hat\Omega\Gamma+\Lambda^{-1}\dot z$, one has $u=\dot\Gamma+\hat\Omega\Gamma$ (cf.~(42) below). Because $p_\mu\dot z^\mu=m\,\eta_{\mu\nu}\Lambda^\nu{}_0\dot z^\mu=m\,\eta_{0\rho}u^\rho\equiv m\,u_0$, the reduced Lagrangian in these variables reads
\begin{equation}
l(\hat\Omega,u)=m\,u_0+\frac{i\lambda}{2}\Tr\big(M_{12}\hat\Omega\big)=\langle\mu,\xi\rangle, \tag{38b}
\end{equation}
which is \emph{linear} in $\xi$, with a constant $\mu\in\mathfrak p^*$. The Euler--Poincar\'e equation~\eqref{eq:EP} on the Poincar\'e algebra therefore reduces to $\ad^*_\xi\mu=0$: the body velocity must lie in the isotropy algebra of $\mu$, while the conserved spatial momentum $\Ad^*_{g^{-1}}\mu$~\cite{MarsdenRatiu} encodes the conserved quantities $p^\mu$ and $J^{\mu\nu}$ found above. For the present $\mu$ (body momentum along the time direction and body spin in the $12$-plane) the isotropy algebra is spanned by body time translations and rotations in the $12$-plane, so that $u\propto e_0$, i.e.\ $\dot z\parallel p$, in agreement with the result above. This is the Lagrangian counterpart of the coadjoint-orbit description~\cite{Souriau}: the motion is confined to the directions generated by the stabilizer of $\mu$, in agreement with the gauge symmetries identified in the constrained Hamiltonian analysis of~\cite{AGGKM22}.
\end{remark}

\subsection{Spinning Charged Particle in an External Field}\label{sec:field}
Now we will show that adding interactions breaks the invariance under the action of $SO(3,1)$ and, correspondingly, a complete reduction is not possible for such a system. Nevertheless, the left-trivialized coordinates can be used to obtain the BMT equations as the equations of motion. We again use the following Lagrangian proposed in~\cite{SkagerstamStern}:
\begin{align}
L&=L_0+eA_\mu\dot z^\mu-\frac\alpha m\,p_\mu\dot z^\mu S_{\alpha\beta}F^{\alpha\beta}\\
&=eA_\mu\dot z^\mu+\frac{i\lambda}{2}\Tr\big(M^{12}\Lambda^{-1}\dot\Lambda\big)+p_\mu\dot z^\mu\Big[1-\frac\alpha m S_{\alpha\beta}F^{\alpha\beta}\Big],
\end{align}
where $e$ is the charge of the particle, $\alpha$ is a constant to be determined, and $F^{\mu\nu}$ is the electromagnetic tensor. The choice of the interaction term is not unique~\cite{SkagerstamStern}; we comment on an alternative at the end of this section.

In order to switch to the body-frame coordinates, we use the following transformations:
\begin{equation}
\begin{aligned}
A^\sigma(z)&=\Lambda^\sigma{}_\mu A'^\mu(\Lambda^{-1}z),\\
F^{\mu\nu}(z)&=\Lambda^\mu{}_\alpha\Lambda^\nu{}_\beta F'^{\alpha\beta}(\Lambda^{-1}z).
\end{aligned}
\end{equation}
A useful identity throughout our calculations will be
\begin{equation}
\dot\Gamma^\mu+\hat\Omega^\mu{}_\rho\Gamma^\rho=\Lambda^{\sigma\mu}\dot z_\sigma.
\end{equation}
With these, the left-trivialized Lagrangian in terms of the body-frame coordinates can be written as
\begin{equation}
\tilde L(\Lambda,\hat\Omega,\Gamma,\dot\Gamma)=\big(\dot\Gamma_0-\hat\Omega^\nu{}_0\Gamma_\nu\big)\underbrace{\big(m-\alpha S'_{\alpha\beta}F'^{\alpha\beta}\big)}_{M(\Phi)}+\frac{i\lambda}{2}\Tr\big(M^{12}\hat\Omega\big)+e\big(\hat\Omega^\mu{}_\nu\Gamma^\nu+\dot\Gamma^\mu\big)A'_\mu(\Gamma),
\end{equation}
where
\begin{equation}
\Phi:=S_{\mu\nu}F^{\mu\nu}(z).
\end{equation}
In this way, we can treat the spin--field interaction term in~(39) merely as a mass redefinition. Physically, $M(\Phi)$ is the rest energy shifted by the magnetic dipole energy: in the rest frame of $p$ one has $S^{0i}=0$ by~(37), and writing $S_{ij}=\epsilon_{ijk}s_k$ and $F_{ij}=\epsilon_{ijk}B_k$, one finds $\Phi=2\,\vec s\cdot\vec B$; with the value $\alpha=eg/4m$ fixed in Section~\ref{sec:BMT}, this gives $M=m-\vec\mu\cdot\vec B$ with $\vec\mu=(ge/2m)\,\vec s$.

Notice that because of the transformation rules~(41), $A'_\mu$ and $F'_{\mu\nu}$ depend on $\Lambda$, so unlike the free case, our Lagrangian has an explicit $\Lambda$ dependency. Therefore, the complete reduction is not possible in this case because of the $\Lambda$ dependency coming from the field terms; in the language of~\eqref{eq:EPgeneral}, the term $T_e^*L_g(\delta\tilde L/\delta g)$ no longer vanishes. As for the heavy top, invariance could be restored by promoting the external field to a variable on which the Lorentz group acts, i.e.\ by treating the field as an advected parameter in the sense of~\cite{HMR98,CHMR98}. For inhomogeneous fields this requires an infinite-dimensional representation space of field configurations; we return to this point in Section~\ref{sec:concl}.

The calculations for the variation of the above action with respect to body coordinates can be found in the Appendix. The resulting equations of motion are
\begin{equation}
\begin{aligned}
0&=\delta z^\sigma\Big(-\dot p_\sigma+e\dot z^\alpha F_{\sigma\alpha}-\alpha\Lambda_{\alpha0}\dot z^\alpha\,\partial_\sigma F^{\mu\nu}S_{\mu\nu}\Big),\\
0&=\frac{i\lambda}{2}[\hat\Omega,\hat\Sigma]^\nu{}_\mu(M^{12})^\mu{}_\nu-\hat\Sigma_{0\mu}\big(\hat\Omega^\mu{}_\nu\Gamma^\nu+\dot\Gamma^\mu\big)M(\Phi)+i\lambda\alpha\big(\dot\Gamma_0-\hat\Omega^\nu{}_0\Gamma_\nu\big)\Tr\big(M^{12}[\Lambda^{-1}F\Lambda,\hat\Sigma]\big),
\end{aligned}
\end{equation}
where
\begin{equation}
p_\mu:=M(\Phi)\Lambda_{\mu0}.
\end{equation}
Identifying $S_{\mu\nu}$ as in~(36), these give us the generalized force and torque equations in~\cite{SkagerstamStern}:
\begin{equation}
\dot p_\sigma=e\,\dot z^\alpha F_{\sigma\alpha}-\alpha\frac{p_\alpha}{M(\Phi)}\dot z^\alpha\,\partial_\sigma F^{\mu\nu}S_{\mu\nu},
\end{equation}
\begin{equation}
\dot S_{\mu\nu}=p_\mu\dot z_\nu-p_\nu\dot z_\mu+2\alpha\frac{p_\sigma\dot z^\sigma}{M(\Phi)}\big(S_{\mu\rho}F^{\rho}{}_{\nu}-S_{\nu\rho}F^{\rho}{}_{\mu}\big).
\end{equation}
This is the torque exerted on a relativistic spinning charged particle in an external field, due to its spin magnetic moment. In the rest frame of $p$ and for $\dot z=p/M$ (to leading order in the field), the field term in~(48) gives, in the notation introduced after~(44), $\dot{\vec s}=\vec\mu\times\vec B$ with $\vec\mu=(ge/2m)\,\vec s$ for the value of $\alpha$ fixed in Section~\ref{sec:BMT}: the torque on a magnetic dipole whose energy $-\vec\mu\cdot\vec B$ appears in $M(\Phi)$. The last term in~(47) is the force on the magnetic moment in an inhomogeneous field (of Stern--Gerlach type).

\subsubsection{Recovering the BMT Equation}\label{sec:BMT}
The Pauli--Lubanski pseudovector is given in terms of our four-momentum and spin tensor by
\begin{equation}
W^\mu=\frac{1}{2m}\epsilon^{\mu\nu\alpha\beta}S_{\alpha\beta}p_\nu,
\end{equation}
which can be inverted by using orthogonality~(37):
\begin{equation}
S_{\gamma\sigma}=\frac1m\,\epsilon_{\mu\rho\gamma\sigma}W^\mu p^\rho.
\end{equation}
To recover BMT, we identify $\alpha=\frac{eg}{4m}$, where $e$ is the charge of the particle and $g$ is the gyromagnetic ratio. If we assume the fields to be weak, such that
\begin{itemize}
\item the fields are taken to be homogeneous,
\item the mass $M(\Phi)$ is approximated by its zero-field value $m$,
\end{itemize}
then~(47) reduces to the Lorentz force
\begin{equation}
\dot p_\mu=eF_{\mu\nu}\dot z^\nu,
\end{equation}
and if we also keep terms only to order $e$,
\begin{equation}
p_\nu=m\dot z_\nu+\mathcal O(e),
\end{equation}
which can be seen by differentiating~(37). Using these, we can write
\begin{equation}
\begin{aligned}
\dot W^\alpha&=\frac{1}{2m}\epsilon^{\alpha\sigma\mu\nu}\big(\dot S_{\mu\nu}p_\sigma+S_{\mu\nu}\dot p_\sigma\big)\\
&=\frac{1}{2m}\epsilon^{\alpha\sigma\mu\nu}\Big(p_\sigma\Big[p_\mu\dot z_\nu-p_\nu\dot z_\mu+2\alpha\frac{p\cdot\dot z}{m}\big(S_{\mu\rho}F^\rho{}_\nu-S_{\nu\rho}F^\rho{}_\mu\big)\Big]+eF_{\sigma\beta}\dot z^\beta S_{\mu\nu}\Big)\\
&=\frac{1}{2m}\epsilon^{\alpha\sigma\mu\nu}\Big(-\frac{eg}{m}p_\sigma S_{\mu\rho}F^\rho{}_\nu+eF_{\sigma\beta}\dot z^\beta S_{\mu\nu}\Big)+\mathcal O(e^2).
\end{aligned}
\end{equation}
Then using~(52) and~(50), we end up with
\begin{equation}
\dot W^\mu=\frac{eg}{2m}F^{\mu\nu}W_\nu+\frac{e}{2m}(g-2)\,\dot z^\mu\dot z_\sigma F^{\sigma\nu}W_\nu+\mathcal O(e^2).
\end{equation}
So we see that, under the weak-field assumption and keeping terms only to linear order in $e$, the torque equation~(48) gives us the Bargmann--Michel--Telegdi (BMT) equation for the spinning charged particle in an external field~\cite{BMT} (see, e.g.,~\cite{Jackson,Barut} for standard derivations). For $g=2$ the second term vanishes; the $(g-2)$ structure reflects the partial cancellation between Larmor and Thomas precession~\cite{Jackson}. Its sign is fixed by the condition $W\cdot\dot z=0$ together with $\dot z_\sigma\dot z^\sigma=-1$; with the signature $(+,-,-,-)$ used in~\cite{Jackson} this term appears with the opposite sign.

As a closing remark, we note that the spin--field interaction term in~(39) is not unique. One such alternative choice is the Lagrangian
\begin{equation}
L_{\rm int}=\alpha\sqrt{-\dot z^2}\,S_{\mu\nu}F^{\mu\nu},
\end{equation}
which is analyzed in~\cite{SkagerstamStern}, and also leads to the BMT equation. This Lagrangian can also be expressed in left-trivialized coordinates; however, the variational method becomes much more complicated due to the $\sqrt{-\dot z^2}$ term. This is not the case for the standard spacetime coordinates, where the variation of~(55) is even simpler than that of~(39). In this sense, the Euler--Poincar\'e framework favours~(39) over its alternative forms on practical grounds: in left-trivialized coordinates the coupling~(39) only rescales the term that is already present in the free Lagrangian, and its variation is straightforward. We stress that this is a statement about computational naturalness within the framework, not a selection principle: both couplings lead to the BMT equation~\cite{SkagerstamStern}. The preference also aligns with physical intuition, as~(39) displays the mass redefinition due to the spin--field coupling explicitly at the level of the Lagrangian.

\section{Conclusion and Outlook}\label{sec:concl}
We have carried out an explicit Lagrangian Euler--Poincar\'e treatment of the relativistic spinning particle on the Poincar\'e group in left-trivialized coordinates. For the free particle the reduction is complete; it parallels the reduction of the heavy top on $SO(3)\ltimes\mathbb R^3$, with the important difference that the spacetime position is a dynamical variable rather than an advected parameter, and the reduced equations are the Lagrangian counterpart of the coadjoint-orbit description (Remark~\ref{rem:full}). External electromagnetic fields reintroduce an explicit dependence on the Lorentz group element, so only a partial reduction is possible; the resulting force and torque equations reproduce the BMT equation for weak homogeneous fields. In left-trivialized coordinates, the spin--field coupling of~\cite{SkagerstamStern} becomes a field-dependent mass.

Two directions seem natural. First, the closer analogue of the direction of gravity is the external field itself: for homogeneous fields the body-frame field $F'=\Lambda^{-1}F\Lambda$ (with $F$ regarded as the matrix $F^\mu{}_\nu$) obeys $\frac{d}{dt}F'=[F',\hat\Omega]$, in direct analogy with~(27b), and the spin--field coupling pairs the body-fixed spin with this advected field, much as the heavy-top potential pairs $\chi$ with $\Gamma$. Treating the field systematically as an advected quantity, in the semidirect-product framework of~\cite{HMR98,CHMR98} or by Lagrangian reduction by stages~\cite{CMR01}, may yield a complete reduction of the interacting system; for inhomogeneous fields this requires an infinite-dimensional space of field configurations. Second, a deeper analysis of such reductions for spinning particles in the generally covariant case may give further insight into the admissible Lagrangians~\cite{BMSS80,BaileyIsrael,KL26}, a problem we plan to investigate in the future.

\section*{Acknowledgements}
We thank Ahmet Ate\c{s} Geni\c{s}y\"urek for valuable discussions.

\appendix
\section{Variational Calculations for the Interacting Lagrangian}
To obtain the equations of motion for a charged particle in an external field, we consider the variation of~(43) with respect to body coordinates:
\begin{equation}\label{eq:A-dL}
\begin{aligned}
\delta\tilde L=\;&\delta\Gamma^\nu\Big(-\dot M(\Phi)\eta_{\nu0}-\hat\Omega_{\nu0}M(\Phi)-e\dot A'_\nu+e\hat\Omega^\mu{}_\nu A'_\mu+e\hat\Omega^\mu{}_\sigma\Gamma^\sigma\frac{\partial A'_\mu}{\partial\Gamma^\nu}+e\dot\Gamma^\mu\frac{\partial A'_\mu}{\partial\Gamma^\nu}\Big)\\
&+\delta\hat\Omega^\nu{}_\mu\Big(\frac{i\lambda}{2}(M^{12})^\mu{}_\nu-\delta^\mu{}_0\Gamma_\nu M(\Phi)+e\Gamma^\mu A'_\nu\Big)\\
&+\big(\dot\Gamma_0-\hat\Omega^\nu{}_0\Gamma_\nu\big)\delta M(\Phi)+e\big(\hat\Omega^\mu{}_\nu\Gamma^\nu+\dot\Gamma^\mu\big)\delta A'_\mu\big|_\Lambda,
\end{aligned}
\end{equation}
where
\begin{equation}\label{eq:A-dM}
\delta M(\Phi)=-\alpha S'_{\alpha\beta}\Big(\frac{\partial F'^{\alpha\beta}}{\partial\Gamma^\nu}\delta\Gamma^\nu+\delta F'^{\alpha\beta}\big|_\Lambda\Big).
\end{equation}
In the last line of~\eqref{eq:A-dL}, we isolated the explicit $\Lambda$ dependencies due to~(41). They can be evaluated using
\begin{equation}\label{eq:A-dA}
\delta A'_\mu\big|_\Lambda=\delta\Lambda^\sigma{}_\mu A_\sigma(z)=\big(\Lambda^\sigma{}_\rho\hat\Sigma^\rho{}_\mu\big)A_\sigma=\hat\Sigma^\rho{}_\mu A'_\rho
\end{equation}
and
\begin{equation}\label{eq:A-dF}
\begin{aligned}
S'_{\alpha\beta}\,\delta F'^{\alpha\beta}\big|_\Lambda&=\delta\big(S'_{\alpha\beta}F'^{\alpha\beta}\big)\big|_\Lambda=-i\lambda\,\delta\Tr\big(M^{12}\Lambda^{-1}F\Lambda\big)\big|_\Lambda\\
&=-i\lambda\Big[\Tr\big(M^{12}(-\Lambda^{-1}\delta\Lambda\,\Lambda^{-1})F\Lambda\big)+\Tr\big(M^{12}\Lambda^{-1}F\,\delta\Lambda\big)\Big]\\
&=-i\lambda\Tr\big(M^{12}[\Lambda^{-1}F\Lambda,\hat\Sigma]\big).
\end{aligned}
\end{equation}
Plugging these, \eqref{eq:A-dL} becomes
\begin{equation}\label{eq:A-dL2}
\begin{aligned}
\delta\tilde L=\;&\delta\Gamma^\nu\Big(-\dot M(\Phi)\eta_{\nu0}-\hat\Omega_{\nu0}M(\Phi)-e\dot A'_\nu+e\hat\Omega^\mu{}_\nu A'_\mu+e\hat\Omega^\mu{}_\sigma\Gamma^\sigma\frac{\partial A'_\mu}{\partial\Gamma^\nu}+e\dot\Gamma^\mu\frac{\partial A'_\mu}{\partial\Gamma^\nu}\\
&\qquad\;-\alpha\big(\dot\Gamma_0-\hat\Omega^\sigma{}_0\Gamma_\sigma\big)S'_{\alpha\beta}\frac{\partial F'^{\alpha\beta}}{\partial\Gamma^\nu}\Big)
+\delta\hat\Omega^\nu{}_\mu\Big(\frac{i\lambda}{2}(M^{12})^\mu{}_\nu-\delta^\mu{}_0\Gamma_\nu M(\Phi)+e\Gamma^\mu A'_\nu\Big)\\
&+i\lambda\alpha\big(\dot\Gamma_0-\hat\Omega^\nu{}_0\Gamma_\nu\big)\Tr\big(M^{12}[\Lambda^{-1}F\Lambda,\hat\Sigma]\big)+e\big(\hat\Omega^\mu{}_\nu\Gamma^\nu+\dot\Gamma^\mu\big)\hat\Sigma^\rho{}_\mu A'_\rho.
\end{aligned}
\end{equation}
Now we can expand the variations according to~(33). We notice that when deriving the expansion for $\delta\Gamma$, the first part $-\hat\Sigma\Gamma$ was due to the $\Lambda$ dependency and the second part $W$ was due to the $z$ dependency. Since we already accounted for the $\Lambda$ variations of $A'_\mu$ and $F'_{\mu\nu}$ explicitly, when expanding $\delta\Gamma$ above, we should only include the $W$ part for the terms containing the partial derivatives of $A'_\mu$ and $F'_{\mu\nu}$. Once again, we set the terms including $W$ and $\hat\Sigma$ separately to $0$, and end up with
\begin{equation}\label{eq:A-W}
\begin{aligned}
0=W^\nu\Big(&-\dot M(\Phi)\eta_{\nu0}-\hat\Omega_{\nu0}M(\Phi)-e\dot A'_\nu+e\hat\Omega^\mu{}_\nu A'_\mu+e\hat\Omega^\mu{}_\sigma\Gamma^\sigma\frac{\partial A'_\mu}{\partial\Gamma^\nu}\\
&+e\dot\Gamma^\mu\frac{\partial A'_\mu}{\partial\Gamma^\nu}-\alpha\big(\dot\Gamma_0-\hat\Omega^\sigma{}_0\Gamma_\sigma\big)S'_{\alpha\beta}\frac{\partial F'^{\alpha\beta}}{\partial\Gamma^\nu}\Big),
\end{aligned}
\end{equation}
\begin{equation}\label{eq:A-Sigma}
\begin{aligned}
0=\;&\big(-\hat\Sigma^\nu{}_\alpha\Gamma^\alpha\big)\Big(-\dot M(\Phi)\eta_{\nu0}-\hat\Omega_{\nu0}M(\Phi)-e\dot A'_\nu+e\hat\Omega^\mu{}_\nu A'_\mu\Big)\\
&+\big(\dot{\hat\Sigma}{}^\nu{}_\mu+[\hat\Omega,\hat\Sigma]^\nu{}_\mu\big)\Big(\frac{i\lambda}{2}(M^{12})^\mu{}_\nu-\delta^\mu{}_0\Gamma_\nu M(\Phi)+e\Gamma^\mu A'_\nu\Big)\\
&+i\lambda\alpha\big(\dot\Gamma_0-\hat\Omega^\nu{}_0\Gamma_\nu\big)\Tr\big(M^{12}[\Lambda^{-1}F\Lambda,\hat\Sigma]\big)+e\big(\hat\Omega^\mu{}_\nu\Gamma^\nu+\dot\Gamma^\mu\big)\hat\Sigma^\rho{}_\mu A'_\rho.
\end{aligned}
\end{equation}

\subsection{The Force Equation}
Plugging $W_\nu=(\Lambda^{-1})_{\nu\sigma}\delta z^\sigma$ and using the identity~(42), equation~\eqref{eq:A-W} simplifies to
\begin{equation}\label{eq:A-force1}
\begin{aligned}
0&=\delta z^\sigma\Big(-\dot M(\Phi)\Lambda_{\sigma0}-M(\Phi)\dot\Lambda_{\sigma0}
+e\underbrace{(\Lambda^{-1})^\nu{}_\sigma\big(\hat\Omega^\mu{}_\nu A'_\mu-\dot A'_\nu\big)}_{-\dot A_\sigma=-\dot z^\alpha\partial_\alpha A_\sigma}
+e\underbrace{(\Lambda^{-1})^\nu{}_\sigma\big(\dot\Gamma^\mu+\hat\Omega^\mu{}_\rho\Gamma^\rho\big)\frac{\partial A'_\mu}{\partial\Gamma^\nu}}_{\dot z^\alpha\partial_\sigma A_\alpha}\\
&\qquad\qquad-\alpha\underbrace{(\Lambda^{-1})^\nu{}_\sigma\big(\dot\Gamma_0-\hat\Omega^\rho{}_0\Gamma_\rho\big)S'^{\alpha\beta}\frac{\partial F'_{\alpha\beta}}{\partial\Gamma^\nu}}_{\Lambda_{\alpha0}\dot z^\alpha\partial_\sigma F^{\mu\nu}S_{\mu\nu}}\Big)\\
&=\delta z^\sigma\Big(-\dot p_\sigma+e\dot z^\alpha F_{\sigma\alpha}-\alpha\Lambda_{\alpha0}\dot z^\alpha\,\partial_\sigma F^{\mu\nu}S_{\mu\nu}\Big),
\end{aligned}
\end{equation}
where
\begin{equation}\label{eq:A-p}
p_\mu:=M(\Phi)\Lambda_{\mu0}.
\end{equation}
This gives us~(47):
\begin{equation}\label{eq:A-force}
\dot p_\sigma=e\,\dot z^\alpha F_{\sigma\alpha}-\alpha\frac{p_\alpha}{M(\Phi)}\dot z^\alpha\,\partial_\sigma F^{\mu\nu}S_{\mu\nu}.
\end{equation}

\subsection{The Torque Equation}
In~\eqref{eq:A-Sigma}, we integrate by parts the terms with $\dot M(\Phi)$ and $\dot A'_\nu$ and use the fact that the term $\dot{\hat\Sigma}{}^\nu{}_\mu(M^{12})^\mu{}_\nu$ is a surface term that vanishes. After cancellations we obtain
\begin{equation}\label{eq:A-torque0}
0=\frac{i\lambda}{2}[\hat\Omega,\hat\Sigma]^\nu{}_\mu(M^{12})^\mu{}_\nu-\hat\Sigma_{0\mu}\big(\hat\Omega^\mu{}_\nu\Gamma^\nu+\dot\Gamma^\mu\big)M(\Phi)+i\lambda\alpha\big(\dot\Gamma_0-\hat\Omega^\nu{}_0\Gamma_\nu\big)\Tr\big(M^{12}[\Lambda^{-1}F\Lambda,\hat\Sigma]\big).
\end{equation}
We will evaluate each term in~\eqref{eq:A-torque0} by expanding $\hat\Sigma$ according to~(35). The first term yields
\begin{equation}\label{eq:A-t1}
\begin{aligned}
\frac{i\lambda}{2}\big(\hat\Omega^\nu{}_\rho\hat\Sigma^\rho{}_\mu-\hat\Omega^\mu{}_\rho\hat\Sigma^\rho{}_\nu\big)(M^{12})^\mu{}_\nu
&=\frac{-\lambda}{2}\Big(\Tr\big(M^{12}\underbrace{\Lambda^{-1}\dot\Lambda\Lambda^{-1}}_{-\frac{d}{dt}(\Lambda^{-1})}M^{\alpha\beta}\Lambda\big)-\Tr\big(M^{12}\Lambda^{-1}M^{\alpha\beta}\dot\Lambda\big)\Big)\theta_{\alpha\beta}\\
&=\dot S^{\alpha\beta}\theta_{\alpha\beta}.
\end{aligned}
\end{equation}
Using~(42), the second term yields
\begin{equation}\label{eq:A-t2}
\begin{aligned}
-\hat\Sigma_{0\mu}(\Lambda^{-1})^\mu{}_\rho\dot z^\rho M(\Phi)&=-\Big(i\theta_{\alpha\beta}\big(\Lambda^{-1}M^{\alpha\beta}\Lambda\big)_{0\mu}\Big)(\Lambda^{-1})^\mu{}_\rho\dot z^\rho M(\Phi)\\
&=-i\,p_\sigma\dot z^\rho(M^{\alpha\beta})^\sigma{}_\rho=\theta_{\alpha\beta}\big(p^\beta\dot z^\alpha-p^\alpha\dot z^\beta\big).
\end{aligned}
\end{equation}
Finally, again using~(42), the third term gives
\begin{equation}\label{eq:A-t3}
\begin{aligned}
&+i\lambda\alpha\Lambda_{\nu0}\dot z^\nu\Tr\big(M^{12}[\Lambda^{-1}F\Lambda,\hat\Sigma]\big)\\
&\quad=-\lambda\alpha\frac{p_\nu\dot z^\nu}{M(\Phi)}\Tr\big(M^{12}\Lambda^{-1}[F,M^{\alpha\beta}]\Lambda\big)\theta_{\alpha\beta}\\
&\quad=-\lambda\alpha\frac{p\cdot\dot z}{M(\Phi)}(M^{12})^\rho{}_\mu(\Lambda^{-1})^\mu{}_\nu\underbrace{[F,M^{\alpha\beta}]^\nu{}_\sigma}_{F^\beta{}_\gamma(M^{\gamma\alpha})^\nu{}_\sigma-F^\alpha{}_\gamma(M^{\gamma\beta})^\nu{}_\sigma}\Lambda^\sigma{}_\rho\,\theta_{\alpha\beta}\\
&\quad=-\lambda\alpha\frac{p\cdot\dot z}{M(\Phi)}\Big(\Tr\big(M^{12}\Lambda^{-1}M^{\gamma\alpha}\Lambda\big)F^\beta{}_\gamma-\Tr\big(M^{12}\Lambda^{-1}M^{\gamma\beta}\Lambda\big)F^\alpha{}_\gamma\Big)\theta_{\alpha\beta}\\
&\quad=-2\alpha\frac{p\cdot\dot z}{M(\Phi)}\Big(S^{\gamma\alpha}F^\beta{}_\gamma-S^{\gamma\beta}F^\alpha{}_\gamma\Big)\theta_{\alpha\beta}.
\end{aligned}
\end{equation}
Combining these results, we obtain~(48):
\begin{equation}\label{eq:A-torque}
\dot S_{\mu\nu}=p_\mu\dot z_\nu-p_\nu\dot z_\mu+2\alpha\frac{p_\sigma\dot z^\sigma}{M(\Phi)}\big(S_{\mu\rho}F^{\rho}{}_{\nu}-S_{\nu\rho}F^{\rho}{}_{\mu}\big).
\end{equation}


\end{document}